\documentclass[aps,pre,twocolumn,amsmath,amssymb,groupedaddress,showpacs]{revtex4}
\usepackage[dvips]{graphicx}
\usepackage{amsmath,amssymb}

\let\leq\leqslant

\usepackage{color}
\usepackage{epsf}
\newcommand{\be}{\begin{equation}}
\newcommand{\ee}{\end{equation}}
\newcommand{\ba}{\begin{array}{l}}
\newcommand{\ea}{\end{array}}
\newcommand{\baa}{\begin{eqnarray}}
\newcommand{\eaa}{\end{eqnarray}}
\newcommand{\lab}[1]{\label{#1}}
\newcommand{\re}[1]{(\ref{#1})}
\newcommand{\ci}[1]{\cite{#1}}

\begin{document}

\title {Nonlinear standing waves on planar branched systems: Shrinking into metric graph.}
\author{Z. Sobirov $^{a}$, D. Babajanov $^b$ and D. Matrasulov $^b$}

\affiliation{$^a$  Tashkent Financial Institute, 60A, Amir Temur Str., 100000, Tashkent, Uzbekistan\\
$^b$ Turin Polytechnic University in Tashkent, 17 Niyazov Str.,
100095,  Tashkent, Uzbekistan}

\email{dmatrasulov@gmail.com}

 % insert PACS

\begin{abstract}
We treat the stationary nonlinear Schr\"odinger equation on two-dimensional branched domains, so-called fat graphs.
The shrinking limit when the domain  becomes one-dimensional metric graph is studied
by using analytical estimate of the convergence of fat graph boundary conditions into those for metric graph.
Detailed analysis of such convergence on the basis of numerical solution of stationary nonlinear Schrodinger equation on a fat graph is provided.
Possibility for reproducing different metric graph boundary conditions studied in earlier works is shown.
Practical applications of the proposed model for such problems as Bose-Einstein condensation in networks, branched optical media, DNA, conducting polymers and wave dynamics in branched capillary networks are discussed.

\end{abstract}
\maketitle

\section{Introduction}
Branched structures and networks appear in many physical systems
and in complex systems from biology, ecology, sociology, economy
and finance \ci{Barabasi1,Havlin}. Particle and wave dynamics in
such systems can be effectively modeled by  nonlinear evolution
equations on metric graphs. The latter are one dimensional system
of bonds which are connected at one or more vertices (branching
points). The connection rule is called the topology of the graph.
When the bonds can be assigned a length, the graph is called a
metric graph. Recently nonlinear evolution equations on metric
graph attracted much attention in the literature (see the Refs.
\ci{zar2010} -\ci{Noja2015}). To solve nonlinear evolution
equations on networks one need to impose boundary conditions on
graph vertices. Soliton solutions providing reflectionless
transmission at the graph vertex together with integrable boundary
conditions were derived in \ci{zar2010}. Different aspects of
nonlinear Schrodinger equation including soliton solutions are
discussed in the Refs.\ci{adami-eur,adami-jpa}. Solutions of
stationary nonlinear Schr\"odinger equation on graphs for
different vertex conditions are obtained in
\ci{adami2011,adami-eur,Karim2013,noja,Noja2015}. Despite the
growing interest to nonlinear evolution equations and soliton
dynamics in networks, most of the studies are still restricted by
considering metric graphs, i.e. by one-dimensional motion wave
motion in branched structures. However, in many cases particle and
wave motion in branched structures   has certain transverse
component, so that the system is two-dimensional. Such systems
should be described within two-dimensional evolution equations on
planar networks.  Such systems can be modeled by so-called fat
graphs. Earlier, the linear Schrodinger equation on fat graphs was
addressed in number of works \ci{expo09}-\ci{expo13} by
considering metric graph limit as transition to from planar to
linear wave motions. Extension of such a study to the case of
nonlinear Schrodinger equation based on the numerical treatment of
the problem was done in recent work \ci{Our2015}.  Fat graph
treatment of sine-Gordon equation on branched Y-junctions was
numerically treated in \ci{caputo14}. Pioneering treatment of
nonlinear evolution equations on fat graphs date back to Kosugi,
who presented in the Ref.\ci{kosugi2002} strict mathematical
treatment of the nonlinear elliptic differential equations on
branched domains and estimates for the shrinking limit. The study
of the nonlinear differential equation on branched domains is
complicated due to the nonlinearity of the boundary conditions
imposed at the branching area or vertex. In case of nonlinear
evolution equations related to physics such boundary conditions
should be derived from conservation laws or other physical
conditions  that greatly simplifies these boundary conditions
\ci{zar2010,Our2015}.

In this work we address the problem of stationary (time-independent) nonlinear Schrodinger equation on fat graphs by focusing on metric graph limit in the shrinking of a fat graph. In particular, we obtain  estimate for the vertex boundary condition in the shrinking limit and show that they reproduce
those for metric graph considered in \ci{adami-eur,Karim2013}. Also, we present a treatment of the problem on the basis of numerical solution of
stationary NLSE on fat graph. For small enough bond widths the numerical solution reproduce metric graph solutions from the Refs.\ci{adami-eur,Karim2013}.

Motivation for the study of stationary NLSE   comes from such
problems as Bose-Einstein condensation on networks
\ci{Barabasi}-\ci{Cast},  standing  waves in branched optical
waveguides \ci{Assanto}, branched liquid crystals and standing
wave-polarons in polymers \ci{Heeger,Heeger1}. All these systems
require taking into account of transverse component of the wave
motion which can be done by considering stationary NLSE on fat
graphs. Estimating of the "boundary" between the planar and linear
motions is of importance for experiments on the study of the wave
motion in nonlinear networks, as such networks are always somehow
planar or tubular.

 This paper is organized as follows. In the next section we give formulation of stationary NLSE on fat and metric graphs. Section III presents the results of numerical solution of stationary NLSE on fat graph at different values of the graphs width by considering shrinking limit. In section IV we present detailed discussion of potential application of fat graph problem to some physical systems having planar branched structure. Finally, section V gives some concluding remarks.

\begin{figure}[ht!]
  \includegraphics[width=80mm]{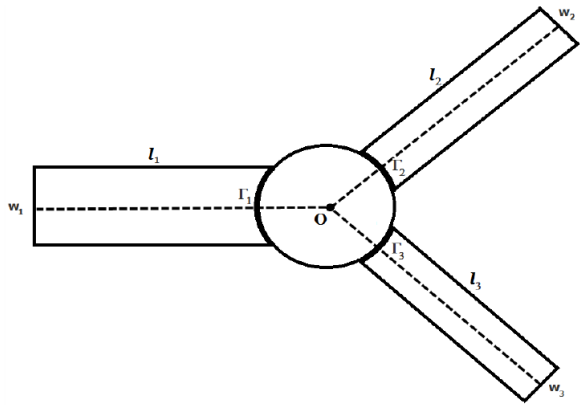}\\[2mm]
\caption{{Sketch of star-shaped fat and metric graphs. Dashed line
presents metric graph. \label{pic1}}}
\end{figure}

\section{Stationary NLSE on a fat graph}

Wave equation on planar branched domains which are often called
"fat graphs" has attracted much attention in the context of wave
dynamics during last decade. In particular, stationary linear wave
equations on fat graphs have been studied in the
Refs.\ci{expo05}-\ci{expo13} (see \ci{Grieser,pobook} and
references therein for detailed reviews).  Corresponding nonlinear
problem is mainly studied for one-dimensional case by considering
metric graph approach Different aspects of the nonlinear
Schrodinger equation on branched one dimensional branched domains
called metric graphs were studied earlier \ci{adami2011}
-\ci{Noja2015}. The metric graph is a set of bonds connected to
each other at the vertices according to a rule which called
topology of a graph with metrics defined in each bond. Recently
the problem of soliton transport in planar branched domains was
studied based on numerical solution of NLSE on fat graph
\ci{Barabasi}. Transition from fat to metric graph problem was
shown in this study.

 Our purpose is to explore solutions of the stationary nonlinear
 Schrodinger equation on two-dimensional branched domain in the limit when the domain shrinks into a metric graph. Therefore we
 introduce two problems, "fat graph" and "metric graph" problems.
Both fat and metric graphs are presented in Fig. 1. Fat graph is a branched domain
having two dimensional bonds and vertex as presented (See Fig.1.).
In the following we denote  bond-lengths of such graph by $l_1, l_2$ and  $l_3$,  bond-widths
by $w_1, w_2$ and $w_3$.  The diameter of the vertex region $\Omega_{\epsilon}=\epsilon\Omega_1$ is $r\epsilon$.
The stationary nonlinear Schrodinger equation (NLSE) on a "fat graph" is given as
\be
\label{nls0}
-\Delta\psi+(V_{\alpha,\epsilon}-\mu)\psi-|\psi|^2\psi=0,
\ee
where the potential
\be
V_{\alpha,\epsilon}(x)=-\frac \alpha {2\pi\epsilon}\exp(-|x|^2/2\epsilon^2)
\lab{pot1}
\ee
is localized  at the vertex domain, $\alpha\in R$ is a parameter. The potential is chosen as to reproduce delta-function in the shrinking limit, i.e.
\be
V_{\alpha,\epsilon} \sim -\alpha\epsilon\delta(x)\;\;\; \text{for $\epsilon\to 0$}.
\ee
\begin{figure}[ht!]
  \includegraphics[width=50mm]{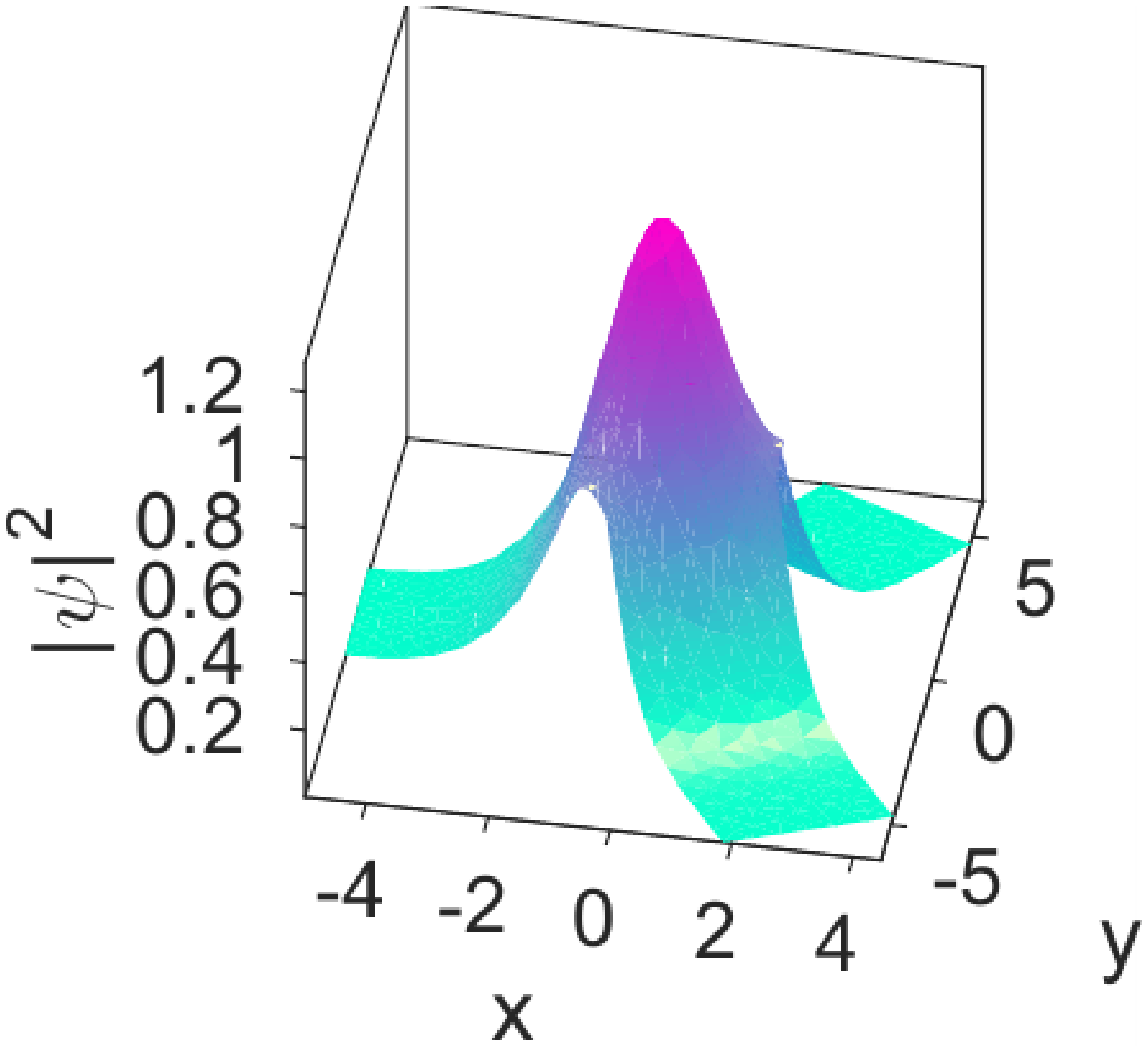}
  \includegraphics[width=50mm]{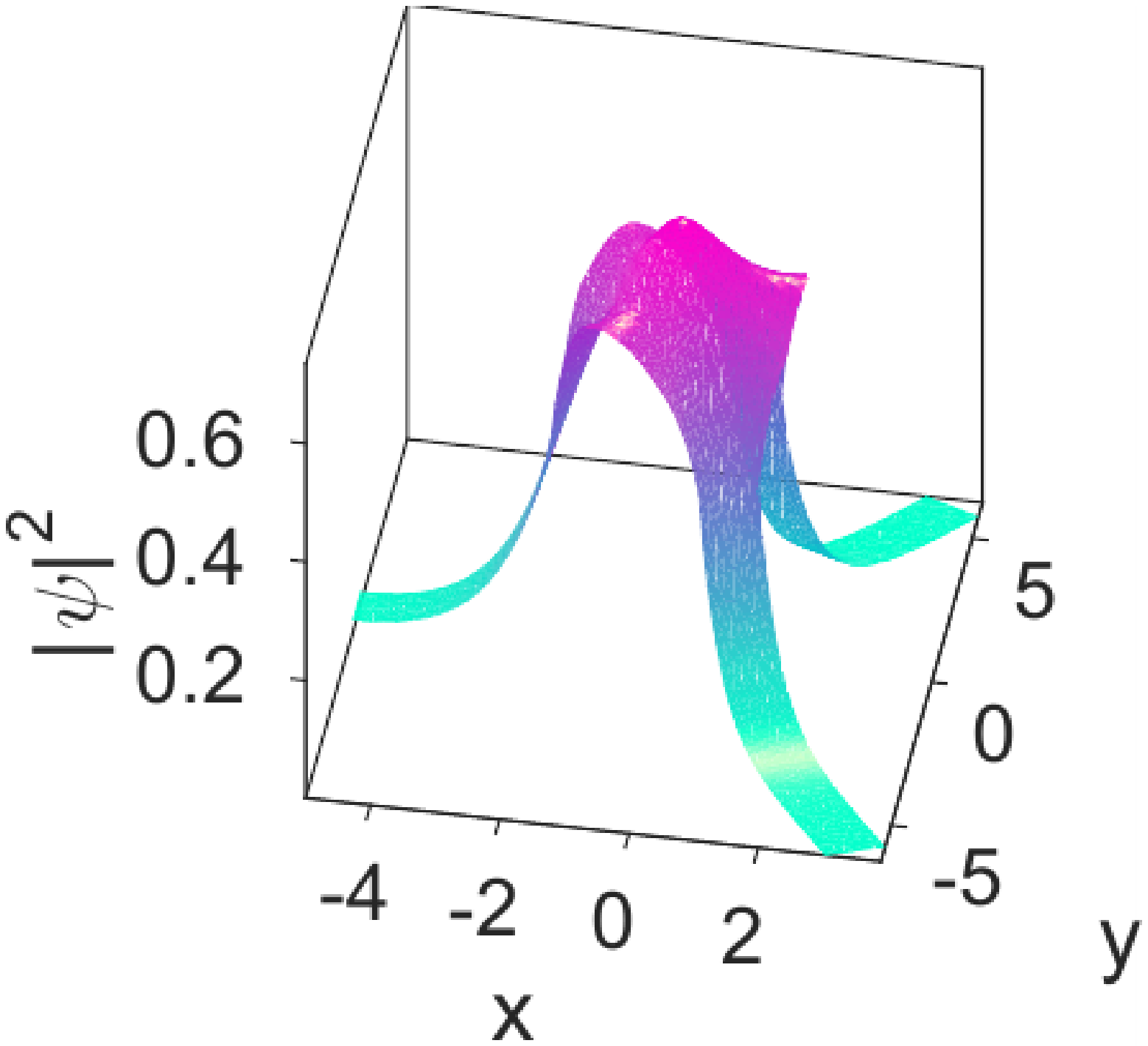}
\caption{{Solution of NLSE on a fat graph at $\epsilon =3$ and
$1,\;\mu =-1, \; \alpha =0.5$ . \label{pic1}}}
\end{figure}
 In the metric graph case we determine coordinate on the $k$th bond from $0$ to $l_k$, $k=1,2,3$.
The metric graph problem is determined by the stationary NLSE which is given on  each bond of the graph as
\be
-\phi''-\mu \phi-|\phi|^2\phi=0.
\lab{nls1}
\ee
The equations are related via the vertex conditions given by
\be
\phi_1=\phi_2=\phi_3,
\lab{vbc01}
\ee
\be
 \phi_1'+\phi_2'+\phi_3'=\alpha\phi_1.
\lab{vbc02}
\ee
In the following the problems given by Eqs.\re{nls1} -\re{vbc02} will be called "metric graph problem".
The aim of this paper is to explore both analytically and numerically the shrinking limit of the fat graph problem given by Eqs.\re{nls0} and \re{pot1}
and determining the conditions providing in the shrinking limit the transition of the  fat graph problem with that of metric graph.
Such an analysis will consist of two parts, analytical estimate for the convergence of the fat graph problem to that
of metric graph and numerical analysis for such convergence. The latter implies the analysis of numerical solution solution of NLSE
for fat graph for different small values of the bond and vertex widths. Here we focus on the
vertex conditions considered in the Refs.\ci{adami-eur,adami2013,noja} which are often called "delta" type boundary conditions.
%\begin{figure}[ht!]
%\includegraphics[width=150mm]{psi1.eps}
%\caption{{Solution of NLSE on a fat graph at $\epsilon =1,\;\mu =-1, \; \alpha =0.5$ . \label{pic1}}}
%\end{figure}

%************************************************************************************************

We divided the fat graph to the following parts: first is vertex region denoted by $\Omega_\epsilon = \epsilon\Omega_1$, where we assumed $\Omega_1$ is
convex region with smooth enough boundary ($0\in \Omega_1$) and tabular (rectangular) parts. In the limit $\epsilon\to 0$ vertex region give as a vertex point $0$,
while tabular parts tends to the bonds of metric graph. Convergence problem in the case of tabular region is well studied.
Here we refer to \cite{kosugi2002}, \cite{jimbo}.
We  focus only vertex region. We denote by $\Gamma_{k,\epsilon}, (k=1,2,3)$ the those parts(arcs)
of the vertex which are connected to the bonds ($\Gamma_{k,\epsilon} =\epsilon\Gamma_k$).
We impose the following boundary conditions for NLSE given by Eq.\re{nls0} in ${\Omega_\epsilon}$ :
\be
\frac{\partial \psi}{\partial n}=0\;\; \text{on}\;\; \partial\Omega_{\varepsilon}/(\Gamma_{1,\varepsilon}\cup\Gamma_{2,\varepsilon}\cup\Gamma_{3,\varepsilon}).
\lab{eq0101}
\ee
Assume that
$$
\frac{\partial \psi}{\partial n}|_{\Gamma_{k,\varepsilon}}\rightarrow\varphi'_k,\;\;
\psi|_{\Gamma_{k,\varepsilon}}\rightarrow\varphi_k\; \text{at}\; \epsilon\rightarrow 0,\ k=1,2,3,
$$
where $\varphi_k$ and $\varphi'_k$ ($k=1,2,3$)are constants.

First we show that $\varphi_1=\varphi_2=\varphi_3$.

According to \cite{jimbo}, \cite{kosugi2002} and maximum principle \cite{Gilbarg}
we have $|\psi(x)|<c_1$,
 $|\nabla \psi(x)|\leq c_2$ for $x\in \Omega_\epsilon$

Denote $v(y)=\psi(y/\epsilon)$.
The function $v(y)$ satisfies the following equation
\be\label{eq_v}
-\Delta_yv(y)-\epsilon^2f(v(y))+\alpha\epsilon V_{1,1}v(y)=0,
\ee
with $f(v)=|v|^2v+\mu v$.
\begin{figure}[ht!]
  \includegraphics[width=80mm]{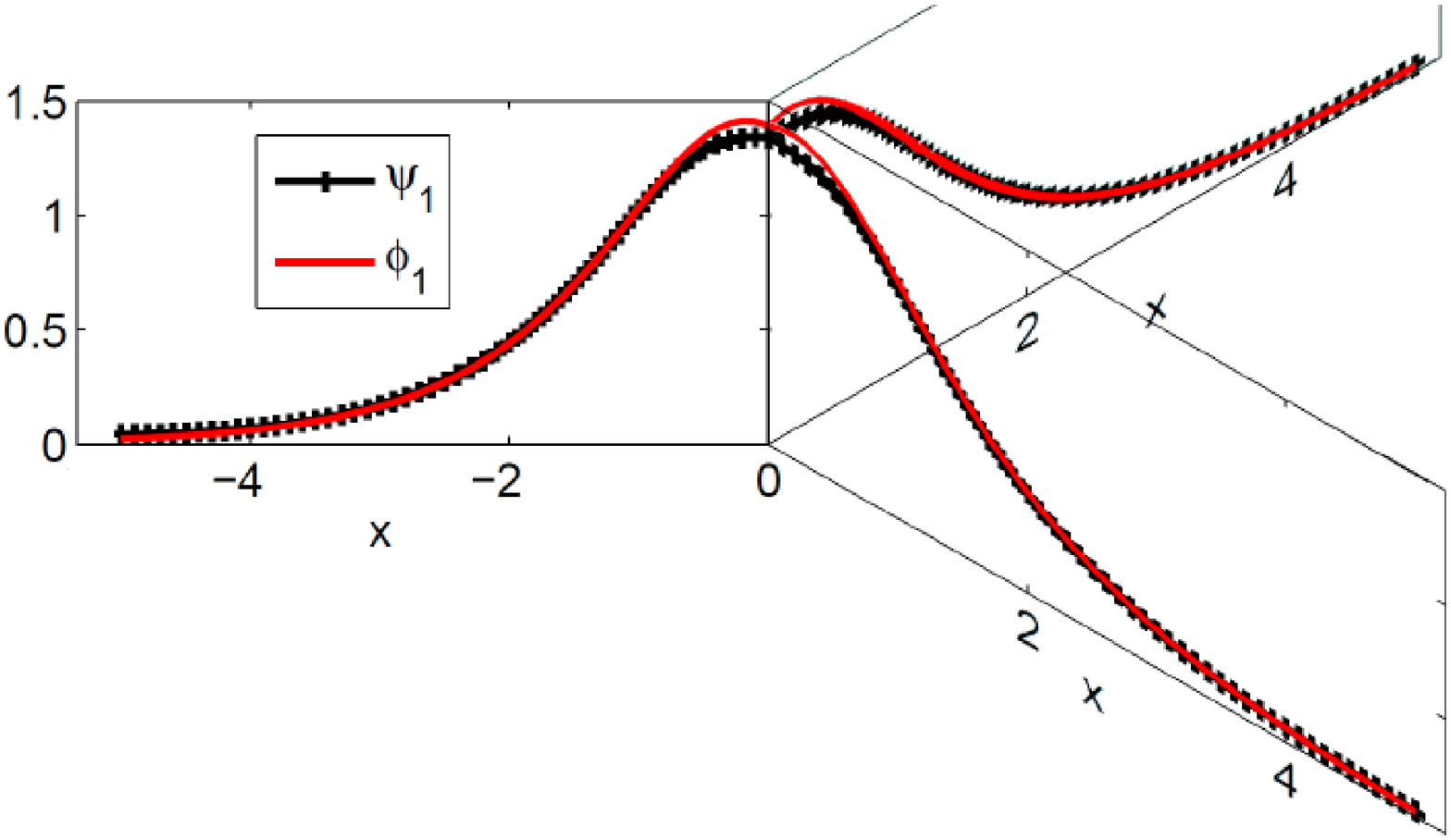}
\caption{{Solution of NLSE on a fat graph at at $\epsilon
=0.5,\;\mu =-1,\; \alpha =0.5$. \label{pic4}}}
\end{figure}
The function $v(y)$ satisfies the estimate (see, e.g.
\cite{kosugi2002}, \cite{Gilbarg}) $\|v\|_{C^2(\Omega_1)}\leq
C_3$. From the Ascoli-Arzel theorem, there exist a sequence
$\epsilon_m\to 0$ and the function $v_{\infty}\in
C^1(\overline{\Omega_1})$, such that $\lim_{k\to\infty}\|v_m-
v_\infty\|_{C^1(\overline{\Omega_1})}=0.$ Where $v_m(y)$ is the
function $v(y)$ with $\epsilon = \epsilon_m$.

Then we have
$$ \iint\limits_{\Omega_1}|\nabla_y v_m(y)|^2dy=
\int\limits_{\partial \Omega_{\epsilon_m}} \psi_m\frac{\partial\psi_m}{\partial n}ds_x-
$$
$$
-\iint\limits{\Omega_{\epsilon_m}} \left[V_{\alpha, \epsilon_m}(x)\psi_m(x)+u_m(x)f(\psi_m(x))\right]dx\leq
$$
\be \leq c_1c_2\epsilon_m\sum_{k=1}^3
|\Gamma_k|+c_1^4|\Omega_1|\epsilon_m^2+c_1\epsilon_m\to 0.
\label{eq9} \ee From Eq.(\re{eq9}) we have $v_\infty=const$, which
proves $\varphi_1=\varphi_2=\varphi_3=\varphi.$

Now we verify the second vertex condition in the limiting problem.
For the mean values of the normal derivatives on the boundary we have ($\epsilon = \epsilon_m$)
\be
\sum\limits_{k=1}^3\frac{1}{\epsilon}\int\limits_{\Gamma_{k,\epsilon}}\frac{\partial
\psi}{\partial n}d\Gamma_{k,\epsilon}\rightarrow\sum\limits_k
|\Gamma_1|\varphi'_k.\lab{eq0101}\ee
On the other hand, integrating Eq.\re{eq0101}
over $\Omega_\epsilon$, we get
\baa
&&\sum\limits_{k=1}^3\frac{1}{\epsilon}\int\limits_{\Gamma_{k,\epsilon}}\frac{\partial
\psi}{\partial
n}d\Gamma_{k,\epsilon}+\alpha\iint\limits_{\Omega_\epsilon}\frac{1}{\epsilon^2}V_{1,1}\left(\frac{x}{\epsilon}\right)\nonumber\\
&-&\frac{1}{\epsilon}\iint\limits_{\Omega_\epsilon}|\psi|^2\psi dx=\frac{\mu}{\epsilon}\iint\limits_{\Omega_\epsilon}\psi dx. \eaa

According to maximum principle \ci{Gilbarg} for small $\epsilon$ we have $\sup|u|\leq C$.
Therefore for the integrals we can write
$$
\frac{1}{\epsilon}\iint\limits_{\Omega_\epsilon}|\psi|^2\psi dx \sim O(\epsilon),\;\;\frac{\mu}{\epsilon}\iint\limits_{\Omega_\epsilon}\psi dx \sim O(\epsilon) ,
$$
for small $\epsilon$. Taking into account the above relations and
properties of the potential $V_{\alpha,\epsilon}$, we obtain \be
\lim_{\epsilon \to
0}\frac{1}{\epsilon}\int\limits_{\Gamma_{k,\epsilon}}\frac{\partial
\psi}{\partial n}d\Gamma_{j,\epsilon}= \alpha \psi(0).
\lab{eq0202} \ee
Eqs. \re{eq0101} and \re{eq0202} lead to \be
\sum\limits_k |\Gamma_k|\varphi'_k=\alpha\varphi. \lab{eq0303} \ee
Thus we showed convergence  of the fat graph problem given by
Eqs.\re{nls0}, \re{pot1} and \re{eq0101} to that for metric graph
with delta type
 boundary conditions given by Eqs.\re{nls1}-\re{vbc02}. In the next section we will show such convergence on the basis of the numerical
 solution of stationary NLSE on fat graph.

%*********************************************************************************************************

\section{Numerical treatment of the shrinking limit}
Our purpose is showing convergence of the stationary NLSE on fat graph into that for metric graph in the shrinking limit using both analytically and numerical analysis of the shrinking limit. Such convergence is of practical importance for various problems dealing the wave dynamics in branched structures where
one needs to neglect by transverse motion of the waves. Here we will focus on the analysis of behavior wave function itself as well as $|\psi(x,y)|^2$. The latter has important physical meaning in the practical applications of the NLSE, e.g., density of particles in BEC, beam intensity in optics, etc.

The estimate presented by  Eqs.\re{eq0202} and \re{eq0303} shows convergence of the vertex boundary conditions of a fat graph into those for metric graph.
Such a convergence can be shown on the basis of numerical the solution of the stationary NLSE on fat graph by considering  the shrinking limit, $\epsilon \to 0$.

Here we will explore numerical solutions of Eq.\re{nls0} with the fat graph boundary conditions given by Eq.\re{eq0101} at different values of the fat graph bond length by considering both attractive and repulsive nonlinearities. They describe bright and dark (static) solitons, respectively.
 Assuming  $w_1 =\epsilon w_2 =\epsilon w_3$, we show that these numerical solutions reproduce the solutions of metric graph problem analytically obtained in the Refs.\ci{adami-eur}.
Solution of the metric graph problem given by Eqs.\re{nls1} -\re{vbc02} was obtained in \ci{adami-eur} and can be written as
\be
\phi_j(x,a) =\sqrt{2\mu}sech(\sqrt{\mu}(x-a^j)),
\lab{metric1}
\ee
where
$$
a^j = \frac{1}{\mu}arctanh\left( \frac{\alpha}{(2j-3)\sqrt{\mu}}\right)\;\; j=1,2,3.
$$
In Fig. 2 solutions of stationary NLSE on fat graph obtained by
numerical solutions of two-dimensional stationary NLSE \re{nls0}
with the boundary conditions \re{eq0101} are for different values
of the bond width  $\epsilon =3, {\rm and} 1$ ($\;\mu =-1,\;
\alpha =0.5$). As it can be seen, the wave function is localized
at the vertex in all cases.

Fig. 3 compares solution of Eq.\re{nls0} for  the bond width,
$\epsilon =1$ with that of for metric graph given by Eqs.\re{nls1}
-\re{vbc02}.  It can be observed the convergence of the solution
of fat graph problem to that of metric graph in the shrinking
limit.

Besides NLSE with attractive (focusing) nonlinearity given by
Eq.\re{nls0}, one can consider repulsive case when the wave
equation given as \be \label{nls001}
-\Delta\psi+(V_{\alpha,\epsilon}-\mu)\psi+|\psi|^2\psi=0. \ee In
Fig.4 solutions for fat and metric graphs are compared for the
case of repulsive nonlinearity and $\epsilon =0.3$. Unlike to
attractive case,  the wave function localized on the bonds for
this repulsive nonlinearity. Fig. 5 shows how the solution NLSE on
fat graph depends on chemical potential $\mu$.
%\begin{figure}[ht!]
%  \includegraphics[width=150mm]{psi2.eps}\\[2mm]
%\caption{{Solution of NLSE on a fat graph at at $\epsilon =0.5,\;\mu =-1,\; \alpha =0.5$. \label{pic1}}}
%\end{figure}
\begin{figure}[ht!]
\includegraphics[width=80mm]{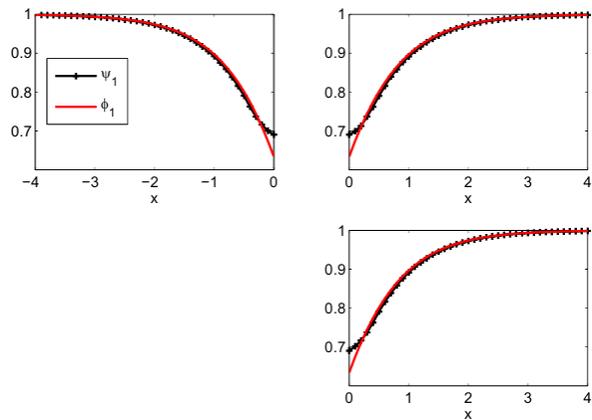}
\caption{{Dark soliton solution of repulsive NLSE on a fat graph
at $\epsilon =0.3,\;\mu =1,\; \alpha =2$. \label{pic10}}}
\end{figure}
\section{Standing nonlinear waves in branched planar structures}
Two-dimensional branched structures where the wave dynamics is described by nonlinear Schrodinger equation appear in many areas of physics. Here we will briefly discuss physical systems where the above model can be applied. \\
\textit{BEC in planar networks}. Bose-Einstein condensation is a
remarkable phenomenon that can be described by a version of NLSE
which is called Gross-Pitaevskii equation
\ci{Barabasi}-\ci{Burioni3}. It can be realized in trapped cold
atoms and depending on the type of a trap NLSE based model has
different potentials and boundary conditions. The above model
describes BEC in planar branched structures/traps which can be
experimentally created in several physical systems. We note that
the Bose-Einstein condensation in networks attracted much
attention recently (see, e.g. Refs.\ci{Barabasi}-\ci{Cast}).
Despite the fact that nonlinear Schrodinger equation is the
powerful tool for description of BEC dynamics, all the studies of
BEC in networks used so far tight binding and statistical
mechanics based approaches \ci{Barabasi}-\ci{Cast}. Moreover, most
of the studies of this problem does not discuss experimental
realization of BEC in networks. Planar BEC can be  experimentally
realized  in surface optical traps \ci{Butov},  superconductive
BEC for exitons in planar systems \ci{Ben} and atom chip films
\ci{Oberthaler}. All these systems can be constructed in in
branched form in which the BEC standing wave can be described by
Eq. \re{nls0}. In this case, parameter $\mu$  in Eq. \re{nls0}
corresponds to chemical potential, while $|\psi|^2$ describes
number of atoms in condensate. Transition from planar to
one-dimensional BEC dynamics corresponding to the shrinking limit
of the above fat graph can be then treated similarly to that in
the Ref.\ci{Carr2000}. Other types of networks and branched
systems where planar and one-dimensional BEC can occur
 are the different types of  Josephson junctions  \ci{Burger}-\ci{Oberthaler}.
 Networks or branched structures of Josephson junctions can be realized in different versions \ci{Hasegawa}-\ci{Luca1}. Planar Josephson junctions can be fabricated in using different techniques \ci{Kras,Nus}.  The standing wave states  of the condensate in planar Josephson network can be described by our model.
Conducting polymers  are also branched systems where BEC can be experimentally realized \ci{Cast}.

%\begin{figure}[ht!]
%  \includegraphics[width=150mm]{psi_4.eps}\\[2mm]
%\caption{{Solution of NLSE on a fat graph at at $\epsilon
%=0.5,\;\mu =-1,\; \alpha =0.5$. \label{pic1}}}
%\end{figure}

\textit{Networks of planar optical waveguides and fibers}. Optical waveguides are the systems where the  wave propagation can be described by linear or nonlinear Schrodinger equation. Stationary nonlinear Schrodinger equation describes standing wave modes in such waveguides. Optical waveguides and fibers can be realized in linear, planar and cylindric forms.
Few papers discuss different ways for  experimental realizations of planar waveguides \ci{Sand,Eis}. Networks of such
waveguides is of importance from the viewpoint of practical importance. In particular, modern optical telecommunication technology requires using networks of such  waveguides rather than separate fibers. Earlier branching of nonlinear optical waves in Y-junction optical media was discussed in the Refs. \ci{Assanto,Santos}. However, these works did not use the above fat or metric graph approaches. In branched optical system, the function $\psi$ in Eq.\re{nls0} describes amplitude of the wave, while quantity $\mu$ corresponds to the propagation constant. By exploring the shrinking limit of NLSE on a fat graph one can determine
minimal width of the fibers for which wave transport in such networks can be considered as one dimensional. Also, the above model can be useful for fabrication
the materials and devices with tunable  optical properties.

\textit{Standing waves in DNA}. Remarkable branched structure
where solitons and nonlinear waves appear is the DNA. Depending on
the model and approach, such waves can be described either by
sine-Gordon or nonlinear Schrodinger equations
\ci{Yakushevich1}-\ci{Sataric}. Within  so-called Peyrard-Bishop
model nonlinear dynamics of DNA base pair is described by NLSE.
The base pair of DNA has a branched structure and can be
considered as a star graph. Realistic wave motion in DNA is
two-dimensional and the transverse component of the oscillations
can play important role in DNA dynamics. Therefore by studying
shrinking limit allows to determine the boundary between the one-
and two-dimensional approaches. Two dimensional model of DNA
dynamics was considered earlier in \ci{Yakushevich1}-\ci{Sataric}.

\textit{Conducting polymers}. Polymers are molecular networks
having complicated topology whose structural units can be often
reduced to star graph. A type of polymers which are called
conjugated polymers can exhibit metallic or semiconductor
properties. Such polymers, which are called conducting polymers
attracted much attention due their wide range of electronic
applications \ci{Heeger, Heeger1}. Charge and spin excitations and
their transport in conducting polymers can have solitons and
standing nonlinear waves described by NLSE. The motion of such
waves is usually two-dimensional that can be effectively described
to one dimensional form. Due to finite size of the branching
points  a polymer chain can be considered as a fat graph. The
above NLSE on fat graph and its shrinking limit ca be powerful
tool for description of charge and spin carriers dynamics in
conducting polymers.  Especially this can be effective method for
charge transport, recombination and separation in polymer based
materials such as light-emitting diodes, organic solar cells etc.

\textit{Capillary networks}. Solitons and nonlinear standing waves
appear in hydrodynamics capillary systems, where fluid dynamics
can be described by NLSE \ci{Rassm}. Planar version of capillary
systems as discussed in\ci{Rassm}. Nonlinear wave in branched
capillary systems should be described by NLSE on a fat graph. The
connection point of such network has finite size. Our model allows
to estimate characteristic width of the capillary tube, which
which wave motion in such network can be considered as linear.

%\begin{figure}[ht!]
%  \includegraphics[width=150mm]{psi_5.eps}\\[2mm]
%\caption{{Solution of NLSE on a fat graph at at $\epsilon
%=0.5,\;\mu =-1,\; \alpha =0.5$. \label{pic1}}}
%\end{figure}
\begin{figure}[ht!]
  \includegraphics[width=80mm]{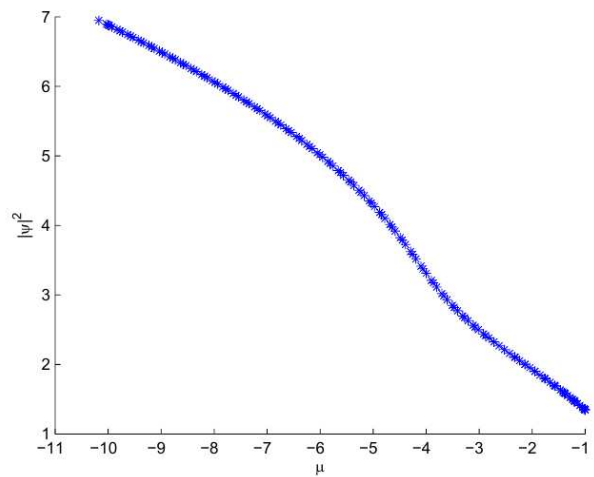}
\caption{{Chemical potential dependence of the solution of NLSE on
a fat graph at the vertex  for $\epsilon =0.5,\;\; \alpha =0.5$.
\label{pic11}}}
\end{figure}

\section{Conclusions}
In this work we studied the stationary nonlinear Schrodinger
equation on fat graphs by focusing on metric graph transition in
the shrinking limit. Analytical estimate for the shrinking of
boundary conditions are obtained. It is shown that in the
shrinking limit fat graph boundary conditions reproduce those for
metric graph. Such a convergence is also shown on the basis of
numerical treatment of NLSE on a fat graph. Detailed discussion of
the potential application of the model to BEC in networks,
branched optical materials, DNA double helix, conducting polymers
and capillary networks is discussed. The model can be extended for
other fat graph topologies, such as tree, ring and complete
graphs. The above study allows to determine the border between
planar and linear motion in branched systems, where particle and
wave transport is effectively considered as one dimensional. The
results of this paper can be useful for the problems of
engineering materials and devices on the basis of branched optical
and electronic structures.

\section*{Acknowledgement} This work is supported by a grant of the Volkswagen Foundation. We thank Hannes Uecker for his valuable comments on the formulation of the problem and for the help in numerical part of the work.

\end{document}